%Paper: hep-th/9305081
%From: Tanmay Vachaspati <TVACHASP@PEARL.TUFTS.EDU>
%Date: Tue, 18 May 1993 16:44 EDT

%%              JNL.TEX                         Doug Eardley
%%
%%      This is a set of TeX 82 macros designed to produce scientific
%%      papers with a minimum of fuss and using as much of plain.tex as
%%      possible.  The user need only know what is in the TeXbook, and
%%      the macros under ``user definitions'' below.  Also, the user
%%      definitions are intended to be as simple as possible, so that
%%      the user may change them as desired.  I have tried to avoid all
%%      cleverness, although it may have snuck in here and there.
%%
%%      For documentation, see the file JNLHLP.TEX.  Optional features are
%%      contained in the files PPT.TEX (for two-up preprints), REFORDER.TEX
%%      (automatic numbering of references), EQNORDER.TEX (automatic numbering
%%      of equations), and TABLEOFC.TEC (automatic generation of table of
%%      contents).

%%
%%  Font definitions suitable for the IMAGEN-480 (Written by Tony Kennedy)
%%      Define a whole menagerie of pseudo-12pt fonts

\font\twelverm=cmr10  scaled 1200   \font\twelvei=cmmi10  scaled 1200
\font\twelvesy=cmsy10 scaled 1200   \font\twelveex=cmex10 scaled 1200
\font\twelvebf=cmbx10 scaled 1200   \font\twelvesl=cmsl10 scaled 1200
\font\twelvett=cmtt10 scaled 1200   \font\twelveit=cmti10 scaled 1200
\font\twelvesc=cmcsc10 scaled 1200
% \font\twelvesf=cmssmc10 scaled 1200
\skewchar\twelvei='177   \skewchar\twelvesy='60

%  Define \...point macros to change fonts and spacings consistently

     %       Melott
\def\twelvepoint{\normalbaselineskip=12.4pt plus 0.1pt minus 0.1pt
  \abovedisplayskip 12.4pt plus 3pt minus 9pt
  \belowdisplayskip 12.4pt plus 3pt minus 9pt
  \abovedisplayshortskip 0pt plus 3pt
  \belowdisplayshortskip 7.2pt plus 3pt minus 4pt
  \smallskipamount=3.6pt plus1.2pt minus1.2pt
  \medskipamount=7.2pt plus2.4pt minus2.4pt
  \bigskipamount=14.4pt plus4.8pt minus4.8pt
  \def\rm{\fam0\twelverm}          \def\it{\fam\itfam\twelveit}%
  \def\sl{\fam\slfam\twelvesl}     \def\bf{\fam\bffam\twelvebf}%
  \def\mit{\fam 1}                 \def\cal{\fam 2}%
  \def\sc{\twelvesc}               \def\tt{\twelvett}
  \def\sf{\twelvesf}
  \textfont0=\twelverm   \scriptfont0=\tenrm   \scriptscriptfont0=\sevenrm
  \textfont1=\twelvei    \scriptfont1=\teni    \scriptscriptfont1=\seveni
  \textfont2=\twelvesy   \scriptfont2=\tensy   \scriptscriptfont2=\sevensy
  \textfont3=\twelveex   \scriptfont3=\twelveex  \scriptscriptfont3=\twelveex
  \textfont\itfam=\twelveit
  \textfont\slfam=\twelvesl
  \textfont\bffam=\twelvebf \scriptfont\bffam=\tenbf
  \scriptscriptfont\bffam=\sevenbf
  \normalbaselines\rm}

%       tenpoint

%%
%%      Various internal macros
%%

\def\beginlinemode{\endmode
  \begingroup\parskip=0pt \obeylines\def\\{\par}\def\endmode{\par\endgroup}}
\def\beginparmode{\endmode
  \begingroup \def\endmode{\par\endgroup}}
\let\endmode=\par
{\obeylines\gdef\
{}}
        %       Melott
        %       Melott
\def\singlespace{\baselineskip=\normalbaselineskip}

\def\oneandahalfspace{\baselineskip=\normalbaselineskip
  \multiply\baselineskip by 3 \divide\baselineskip by 2}
\def\doublespace{\baselineskip=\normalbaselineskip \multiply\baselineskip by 2}

\newcount\firstpageno
\firstpageno=2
%% FOLLOWING LINE CANNOT BE BROKEN BEFORE 80 CHAR
\footline={\ifnum\pageno<\firstpageno{\hfil}\else{\hfil\twelverm\folio\hfil}\fi}

\def\toppageno{\global\footline={\hfil}\global\headline
  ={\ifnum\pageno<\firstpageno{\hfil}\else{\hfil\twelverm\folio\hfil}\fi}}
\let\rawfootnote=\footnote              % We must set the footnote style
\def\footnote#1#2{{\rm\singlespace\parindent=0pt\parskip=0pt
  \rawfootnote{#1}{#2\hfill\vrule height 0pt depth 6pt width 0pt}}}
\def\raggedcenter{\leftskip=4em plus 12em \rightskip=\leftskip
  \parindent=0pt \parfillskip=0pt \spaceskip=.3333em \xspaceskip=.5em
  \pretolerance=9999 \tolerance=9999
  \hyphenpenalty=9999 \exhyphenpenalty=9999 }
\def\dateline{\rightline{\ifcase\month\or
  January\or February\or March\or April\or May\or June\or
  July\or August\or September\or October\or November\or December\fi
  \space\number\year}}
\def\received{\vskip 3pt plus 0.2fill
 \centerline{\sl (Received\space\ifcase\month\or
  January\or February\or March\or April\or May\or June\or
  July\or August\or September\or October\or November\or December\fi
  \qquad, \number\year)}}

%%
%%      Page layout, margins, font and spacing        (feel free to change)
%%

\hsize=6.5truein
\hoffset=0.0truein
\vsize=8.5truein
\voffset=0.25truein
\parskip=\medskipamount
%\toppageno                            % REMOVE FOR NO PAGE NUMBERS. TV.
\twelvepoint
\doublespace
\def\\{\cr}
\overfullrule=0pt % delete the nasty little black boxes for overfull box

%%
%%      The user definitions for major parts of a paper (feel free to change)
%%

  %  "Draft", Timestamp

 % Preprint number at upper right of title page

\def\title#1{                   %  Title on title page
   \null \vskip 3pt plus 0.3fill \beginlinemode
   \doublespace \raggedcenter {\bf #1} \vskip 3pt plus 0.1 fill}

\def\author                     %  Author(s) name(s)  on title page
  {\vskip 3pt plus 0.1fill \beginlinemode \doublespace \raggedcenter}

\def\affil                      % Affiliations (can intermix with \author)
  {\vskip 3pt \beginlinemode \doublespace \raggedcenter \it}

\def\abstract                   % Begin abstract
  {\vskip 3pt plus 0.1fill \subhead {Abstract:}
   \beginparmode \narrower \oneandahalfspace }

\def\endtopmatter               % End title page, begin body of paper
  {\vskip 3pt plus 0.1fill \endpage \body}

\def\body                       % Begin text body;  can be used to end
  {\beginparmode}               % \title, \author, \affil, \abstract,
                                % \reference, or \figurecaption modes

\def\head#1{                    % Head;  NOTE enclose the text in {}
   \goodbreak \vskip 0.4truein  %  e.g., \head{I. Introduction}
  {\immediate\write16{#1} \raggedcenter {\sc #1} \par}
   \nobreak \vskip 3pt \nobreak}

\def\subhead#1{                 % Subhead;  NOTE enclose the text in {}
  \vskip 0.25truein             % e.g., \subhead{A. History of the Problem}
  {\raggedcenter {\it #1} \par} \nobreak \vskip 3pt \nobreak}

\def\beneathrel#1\under#2{\mathrel{\mathop{#2}\limits_{#1}}}

\def\refto#1{${\,}^{#1}$}       % For references in text as superscript

\newdimen\refskip \refskip=0pt
\def\references         % Begin references -- basic format is Ap. J., i.e.
  {\head{References}    %  [journal], [volume], [page] (space after comma).
   \beginparmode \frenchspacing \parindent=0pt \leftskip=\refskip
   \parskip=0pt \everypar{\hangindent=20pt\hangafter=1}}

\gdef\refis#1{\item{#1.\ }}                     % Ref list numbers.

\gdef\journal#1, #2, #3 {               % references,  Ap. J.  style
    {\it #1}, {\bf #2}, #3.}            % comma separates: name, vol, page

      % NB: year BEFORE journal name
%  ex:  J. Smith.  1901.  Gen. Rel. Grav.  123: 456.

% journal abbreviations:
% convention: lower case ("\prd") is Phys. Rev. style

% uppercase ("\ApJ") for Ap. J. style

\def\endreferences{\body}

\def\figurecaptions             % Begin figure captions
  {\endpage \beginparmode \head{Figure Captions}
   \parskip=3pt \everypar{\hangindent=20pt\hangafter=1} }

\def\endpage                    %  Eject a page
  {\vfill\eject}

% Ways to say goodbye
\def\endpaper   {\endmode\vfill\supereject}
\def\endjnl     {\endpaper\end}

%%
%%      Various little user definitions
%%

\def\ref#1{ref.{#1}}                    %   For inline references
\def\Ref#1{Ref.{#1}}                    %       ditto
\def\[#1]{[\cite{#1}]}
\def\cite#1{{#1}}
          %   For citation of equation numbers
        %       ditto
                       %       ditto

                     %       ditto

%\def\Fig{Figure}
%\def\Figs{Figures}
                %   PRL figure caption page
          %   Ap. J. Figure caption page
\def\(#1){(\call{#1})}
\def\call#1{{#1}}
\def\frac#1#2{{#1 \over #2}}
\def\half{  {\frac 12}}

\def\fourth{{\frac 14}}
\def\12{{1\over2}}

\def\sla{\raise.15ex\hbox{$/$}\kern-.57em}
\def\leaderfill{\leaders\hbox to 1em{\hss.\hss}\hfill}
\def\twiddle{\lower.9ex\rlap{$\kern-.1em\scriptstyle\sim$}}
\def\bigtwiddle{\lower1.ex\rlap{$\sim$}}
\def\gtwid{\mathrel{\raise.3ex\hbox{$>$\kern-.75em\lower1ex\hbox{$\sim$}}}}
\def\ltwid{\mathrel{\raise.3ex\hbox{$<$\kern-.75em\lower1ex\hbox{$\sim$}}}}
\def\square{\kern1pt\vbox{\hrule height 1.2pt\hbox{\vrule width 1.2pt\hskip 3pt
   \vbox{\vskip 6pt}\hskip 3pt\vrule width 0.6pt}\hrule height 0.6pt}\kern1pt}
\def\tdot#1{\mathord{\mathop{#1}\limits^{\kern2pt\ldots}}}

\def\pmb#1{\setbox0=\hbox{#1}%
  \kern-.025em\copy0\kern-\wd0
  \kern  .05em\copy0\kern-\wd0
  \kern-.025em\raise.0433em\box0 }

\catcode`@=11
\newcount\r@fcount \r@fcount=0
\newcount\r@fcurr
\immediate\newwrite\reffile
\newif\ifr@ffile\r@ffilefalse
\def\w@rnwrite#1{\ifr@ffile\immediate\write\reffile{#1}\fi\message{#1}}

\def\writer@f#1>>{}
\def\referencefile{%                      Stuff to write .REF file
  \r@ffiletrue\immediate\openout\reffile=\jobname.ref%
  \def\writer@f##1>>{\ifr@ffile\immediate\write\reffile%
    {\noexpand\refis{##1} = \csname r@fnum##1\endcsname = %
     \expandafter\expandafter\expandafter\strip@t\expandafter%
     \meaning\csname r@ftext\csname r@fnum##1\endcsname\endcsname}\fi}%
  \def\strip@t##1>>{}}

\def\citeall#1{\xdef#1##1{#1{\noexpand\cite{##1}}}}
\def\cite#1{\each@rg\citer@nge{#1}}     % Variable No. of args, separated by
%%","

\def\each@rg#1#2{{\let\thecsname=#1\expandafter\first@rg#2,\end,}}
\def\first@rg#1,{\thecsname{#1}\apply@rg}       % each@ag is a general purpose
\def\apply@rg#1,{\ifx\end#1\let\next=\relax%      variable no. of arg. macro.
\else,\thecsname{#1}\let\next=\apply@rg\fi\next}% args separated by commas

\def\citer@nge#1{\citedor@nge#1-\end-}  % Check for M-N range (M and N numbers)
\def\citer@ngeat#1\end-{#1}
\def\citedor@nge#1-#2-{\ifx\end#2\r@featspace#1 % Single argument
  \else\citel@@p{#1}{#2}\citer@ngeat\fi}        % M-N range of arguments
\def\citel@@p#1#2{\ifnum#1>#2{\errmessage{Reference range #1-#2\space is bad.}
    \errhelp{If you cite a series of references by the notation M-N, then M and
    N must be integers, and N must be greater than or equal to M.}}\else%
 {\count0=#1\count1=#2\advance\count1
by1\relax\expandafter\r@fcite\the\count0,%

  \loop\advance\count0 by1\relax%         Loop from M to N
    \ifnum\count0<\count1,\expandafter\r@fcite\the\count0,%
  \repeat}\fi}

\def\r@featspace#1#2 {\r@fcite#1#2,}    % Eat spaces at beginning or end of arg
\def\r@fcite#1,{\ifuncit@d{#1}          % Cite individual reference
    \expandafter\gdef\csname r@ftext\number\r@fcount\endcsname%
    {\message{Reference #1 to be supplied.}\writer@f#1>>#1 to be supplied.\par
     }\fi%
  \csname r@fnum#1\endcsname}

\def\ifuncit@d#1{\expandafter\ifx\csname r@fnum#1\endcsname\relax%
\global\advance\r@fcount by1%
\expandafter\xdef\csname r@fnum#1\endcsname{\number\r@fcount}}

\let\r@fis=\refis                       % Save old \refis, redefine
\def\refis#1#2#3\par{\ifuncit@d{#1}%      Use two params #2 #3 to strip blank
    \w@rnwrite{Reference #1=\number\r@fcount\space is not cited up to now.}\fi%
  \expandafter\gdef\csname r@ftext\csname r@fnum#1\endcsname\endcsname%
  {\writer@f#1>>#2#3\par}}

\def\r@ferr{\endreferences\errmessage{I was expecting to see
\noexpand\endreferences before now;  I have inserted it here.}}
\let\r@ferences=\references
\def\references{\r@ferences\def\endmode{\r@ferr\par\endgroup}}

\let\endr@ferences=\endreferences
\def\endreferences{\r@fcurr=0%            Save old \endreferences, redefine
  {\loop\ifnum\r@fcurr<\r@fcount%         Loop over refnum and produce text
    \advance\r@fcurr by 1\relax\expandafter\r@fis\expandafter{\number\r@fcurr}%
    \csname r@ftext\number\r@fcurr\endcsname%
  \repeat}\gdef\r@ferr{}\endr@ferences}

% Save old \endpaper, redefine it to write parting message.

\let\r@fend=\endpaper\gdef\endpaper{\ifr@ffile
\immediate\write16{Cross References written on []\jobname.REF.}\fi\r@fend}

\catcode`@=12

\citeall\refto          % These macros will generate citations
\citeall\ref            %
\citeall\Ref            %

%%%%%%%%%%%%%%%%%%%%%%%%%%%%%%%%%%%%%%%%%%%%%%%%%%%%%%%%%%%%%%%%%%

\vglue 1. truein
\title
{
Do Global String Loops Collapse to Form Black Holes?
}
\author
{
Joaquim ${\rm Fort}^*$
and
Tanmay Vachaspati
}
\affil
{
Tufts Institute of Cosmology, Department of Physics and Astronomy,
Tufts University, Medford, MA 02155.
}

\abstract
\doublespace

Hawking has shown that the emission of gravitational radiation
cannot prevent circular loops of gauged cosmic strings from collapsing
into black holes. Here we consider the corresponding question for
global strings: can Goldstone boson
emission prevent circular loops of global cosmic strings from forming
black holes? Our results show that for every value of the string
tension there is a certain critical size
below which the circular loop does not collapse to form a black hole.
For GUT scale strings, this
critical size is much larger than the current horizon.

\endtopmatter

Some years ago, Hawking\refto{sh} proved that a circular loop of
gauge cosmic string would eventually collapse to form a black hole.
Furthermore, he showed that in the process of collapse, the loop
would radiate at most 29\% of its total energy in gravitational
radiation before forming a black hole. The analysis Hawking gave
used the singularity theorems of gravity and did
not depend on the field theoretic details of the gauge string.

In the present paper, we will consider a circular loop
of {\it global} string. Here the loop primarily radiates Goldstone
bosons during its collapse and the question arises if a black hole
can eventually form.
It is unfortunate that there are no corresponding singularity theorems
that can give information about non-gravitational radiation and so
we have to use a method which is much less elegant than that
used by Hawking: we explicitly find the energy lost in Goldstone
bosons as a function of time and check if the loop ever collapses to
within its own Schwarzchild radius. If it does, then
a black hole will form while if the loop
never falls within
its Schwarzchild radius then a black hole will not form.

In order to find the energy lost by the loop into Goldstone boson emission,
we have had to make a number of simplifying assumptions.
For example, we have ignored the radiation back-reaction\refto{adjq, ps},
the self-gravity of the loop and
the energy lost to gravitational radiation.
Because of these assumptions, our results cannot be considered
rigorous. However, we feel that a substantially more
complicated calculation without these simplifications would
yield qualitatively similar results.

The simplest field theoretic action that gives global strings is:
$$
S = \int d^4 x \left [ \half |\partial _\mu \phi |^2 -
                       \fourth \lambda ( |\phi |^2 - \eta ^2 )^2
               \right ]
\eqno (1)
$$
where, $\phi$ is a complex scalar field. The dynamics of global string
loops also follows from (1) but it seems that this can only be done
numerically\refto{rdps, ps}. Instead a somewhat different approach is
usually taken and the Kalb-Ramond action
is considered\refto{ew, avtv}:
$$
S = {1 \over 6} \int F_{\mu \nu \sigma} F^{\mu \nu \sigma} +
       2 \pi \eta \int A_{\mu \nu} d \sigma^{\mu \nu} -
        \mu_0 \int d^2 \sigma
\eqno (2)
$$
where, $A_{\mu \nu}$ is an antisymmetric tensor field,
$$
F_{\mu \nu \sigma} = \partial _\mu A_{\nu \sigma} +
                      \partial _\nu A_{\sigma \mu} +
                       \partial _\sigma A_{\mu \nu}
\eqno (3)
$$
and,
the surface element of the string world sheet $x^\mu ( \zeta , \tau )$ is
$$
d\sigma ^{\mu \nu} =
                  c^{\mu \nu } (\zeta , \tau ) d\zeta d\tau \ .
\eqno (4)
$$
Here, $\zeta$ and $\tau$ parametrize the world-sheet and
$$
c^{\mu \nu} = {\dot x}^\mu {x '}^\nu - {\dot x}^\nu {x'}^\mu \  .
\eqno (5)
$$
Overdots and primes denote differentiation with respect to the time
coordinate and $\zeta$ respectively.
In addition,
$$
d\sigma = (- \half d\sigma _{\mu \nu} d\sigma ^{\mu \nu} )^{1/2}
\eqno(6)
$$

The connection between (1) and (2) is
established by the relation\refto{avtv}
$$
{1 \over 6} \epsilon_{\mu \nu \sigma \tau} F^{\nu \sigma \tau}
 = \eta \partial_\mu \theta
\eqno (7)
$$
where, $\theta$ is the phase of the complex scalar field $\phi$.
One can also attempt to derive\refto{rdps} (2) from (1) under
suitable assumptions.
Our attitude in the present work will be to simply adopt (2) as our
starting point. This point
of view is fully justified in the context of cosmic
superstrings\refto{ew} as they are based precisely on the action
in (2).

We now use the gauge choice
$$
\partial_\nu A^{\mu \nu} = 0, \  {\dot x} \cdot { x'} = 0,
\  {\dot x}^2 + { x'}^2 = 0 , \  \tau = t
\eqno (8)
$$
Then the equations of motion following from (2) are:
$$
\partial_\sigma \partial^\sigma A^{\mu \nu} = 4 \pi j^{\mu \nu}
\eqno (9)
$$
$$
j^{\mu \nu} = \half \eta \int d\zeta
               \delta^3 [ \vec x - \vec x (\zeta , t) ]
               c^{\mu \nu} (\zeta , t)
\eqno (10)
$$
$$
\mu_0 ( {\ddot x}_\mu - {x ''}_\mu ) =
        4 \pi \eta F_{\mu \nu \sigma} {\dot x}^\nu {x '}^\sigma \  .
\eqno (11)
$$

The right-hand side of (11) gives the back reaction of the radiation
on the dynamics of the string. It can be
shown\refto{flregge, adjq} that it also
contains a term that renormalizes the bare string tension
$\mu_0 \approx \eta^2 $.
If we ignore the radiation back reaction, the string dynamics
is simply that of a Nambu-Goto string. For a circular loop of
radius $R(t)$, the solution is:
$$
R(t) = R_0 cos( t/R_0 )
\eqno (12)
$$
where, $R_0$ is the radius of the loop at time $t=0$.

We now turn to the radiation from the circular loop. For this we must
find the solution to (9) as a function of time.
This is easily done by standard methods\refto{jackson} and after
using (10) we find,
$$
A^{\mu \nu} (\vec x , t)  =
    \eta \int_0 ^{2\pi} d\zeta \int d \tau \  c^{\mu \nu} (\zeta , \tau )
\  \Theta ( t - \tau ) \  \delta [ (x - x(\zeta , \tau ))^2 ]
\eqno (13)
$$
For a given value of $\zeta$, the integrand over $\tau$ will be non-zero
only when $\tau$ equals the retarded time, $ t_r$, which is defined by,
$$
t_r = t - | \vec x - \vec x (\zeta , t_r )| \  .
\eqno (14)
$$

Differentiating eq. (13) and then using (3) gives a very
lengthy expression for the field strength. However, we are only interested
in the radiation part of the field strength. This means that we should let
$r = |\vec x | \rightarrow \infty$ and keep only the leading $1/r$ terms
of the field strength. This procedure yields,
$$
F^{\mu \nu \sigma} _{rad} = {\eta \over {2r}} \int d\zeta
               \biggl [
    {   {{\dot c}^{\nu \sigma} n^{\mu} + {\dot c}^{\sigma \mu} n^{\nu}
        +{\dot c}^{\mu \nu} n^\sigma } \over
        {[ n^\lambda {\dot x}_\lambda ]^2}
     } -
     {  {c^{\nu \sigma} n^\mu + c^{\sigma \mu} n^\nu + c^{\mu \nu} n^\sigma}
         \over {[n^\lambda {\dot x}_\lambda ] ^3 }
         n^\gamma {\ddot x}_\gamma
     }
                \biggr ] _{t=t_r}
\eqno (15)
$$
where,
$$
n^\mu = {{x^\mu - x^\mu ( \zeta , t_r )} \over
         {|{\vec x} - {\vec x} (\zeta , t_r )| }
        }
\eqno (16)
$$
is a null vector.

We are interested in the flux of energy radiated from the string. This can
be found from the energy-momentum tensor\refto{avtv}:
$$
T^{0i} _{rad} = - F_{i \alpha \beta} ^{(rad)}
                  F^{0 \alpha \beta} _{(rad)} ,
                \ \ \ \  (i \ne 0) \ .
\eqno(17)
$$
The energy radiated from the string is given by an integral of the energy
flux over a sphere of radius $r$ (which is taken to $\infty$):
$$
\dot E = r^2 \int_0 ^\pi d\theta \int _0 ^{2\pi} d\phi \  sin\theta
\   e_i T^{0i} _{(rad)}
\eqno (18)
$$
where, $e_i$ is the unit radial three vector
$$
e_i = (sin\theta cos \phi , sin\theta sin\phi , cos \theta ) \  .
\eqno (19)
$$
Putting together eqs. (15)-(19), we get,
$$
\eqalign{
\dot E = {{\eta^2} \over 2} \int _0 ^\pi d\theta sin\theta cos^2 \theta
  \int_0 ^{2\pi} d\phi
\biggl [
   \int _0 ^{2\pi} d\sigma
\biggl \{
&
   {   { {\dot R}^2 + R {\ddot R} } \over
       {[1-{\dot R} sin\theta cos(\sigma - \phi )]^2}
   } +
\cr
&
   {   {R {\dot R} {\ddot R} sin\theta cos(\sigma - \phi )} \over
       {[1-{\dot R} sin\theta cos(\sigma - \phi )]^3}
   }
\biggr \}
\biggr ]^2
\cr
}
\eqno (20)
$$
where, $R$ and its time derivatives are evaluated at the retarded time
and $\sigma = \zeta /R_0$. Note that the retarded time in the radiation
zone ($r \rightarrow \infty$) is given by $t_r = t - r$.
Therefore, the effect of having the retarded time in (20) is simply
to shift $t$ and this shift may be abosrbed by redefining $t$.
The overall effect is equivalent to evaluating the integrand in (20)
at time $t$ and not at the retarded time $t_r$.

The integrations over $\sigma$ and $\phi$ can now be done to yield
$$
\dot E = \eta^2 \pi^3 \int_0 ^\pi d\theta sin\theta cos^2 \theta
      \biggl [
    {  {2 {\dot R}^2 + 2 R {\ddot R}} \over
        {(1- {\dot R}^2 sin^2 \theta )^{3/2}}
    } +
    {  {3 R {\dot R}^2 {\ddot R} sin^2 \theta} \over
        {(1- {\dot R}^2 sin^2 \theta )^{5/2} }
    }
      \biggr ]^2 \  .
\eqno (21)
$$
The integration over $\theta$ can be done by transforming the
variable of integration to $u = cos\theta$:
$$
\dot E (t) = \eta^2 \pi^3 \left [
{{88 x + 16 x cos(4 x ) + 12 sin(4x) - 19 sin(8x)} \over
{ 512 \  sin^3 x \   cos^3 x }} \right ]
\eqno (22)
$$
where, $x = t/R_0$.

Next we need to find $E(t)$. For this we need to integrate (22) over
$t$. We have done this integration numerically and the result is
shown in Fig. 1.

Our criterion for black hole formation is:
$$
{{2 G M(t)} \over {R(t)}} \ge 1  \  ,
\eqno (23)
$$
for any time $t$.
Here $M(t)$ is the energy of the loop at time $t$, that is,
$$
M(t) = 2\pi R_0 \mu - E(t)  \    .
\eqno (24)
$$
Note that the initial energy, $2\pi R_0 \mu$,
is given in terms of the {\it renormalized}
string tension,
$$
\mu \approx \mu_0 {\rm ln} (\eta R_0) \equiv \eta^2 \Lambda \    ,
\eqno (25)
$$
where, $\Lambda$ varies logarithmically with $R_0$.

Let us define a function $f(t)$ via,
$$
4\pi G \mu f(t) = {{2 G M(t)} \over {R(t)}} \  .
\eqno (26)
$$
Using (12), (24), (25) and (26) we find:
$$
f(t) = {1 \over {cos(t/R_0 )}}
         \left [ 1 - {{E(t)} \over {2\pi R_0 \eta^2 \Lambda}} \right ]
\eqno (27)
$$
The criterion for black hole formation now is,
$
4\pi G \mu > {1 \over {f(t)}}
$
for some $t$.

In Fig. 2, we plot $f(t)$ vs. $x = t/R_0$ for values of $\Lambda$
between 1 and 100. The behaviour of the plots is easily
understandable in terms of two effects present in eq. (27):
(i) the factor of $cos(t/R_0)$ in the denominator - that is,
the collapse of the loop - which tends to increase $f(t)$, and,
(ii) the term $E(t)$ - that is, the energy lost to radiation -
which tends to decrease $f(t)$. For small values of $\Lambda$,
the effect of the radiation is very strong and the decrease
in $f(t)$ due to the rapid increase in $E(t)$ cannot be overcome
by the effects of loop collapse. As a result, $f(t)$ continues
to decrease from $t=0$ until it vanishes. At this point,
the loop has radiated away all its energy. (Realistically, our
calculation breaks down for such small $\Lambda$ since the
radiation is very intense and back-reaction effects will be
important.) When $\Lambda$ is large, the collapse of the loop
is the dominant effect on the behaviour of $f(t)$ and hence
$f(t)$ grows. This growth can only continue for a while, however,
since $E(t)$ is a growing function that blows up at
$t = R_0\pi /2 $. Therefore, $f(t)$ grows for a while, then turns
around and starts decreasing. This shows that $f(t)$ always has
a maximum value, $f_{max}$.

The criterion for black hole formation can now be written as:
$$
4\pi G \mu > f_{max}^{-1}  \  ,
\eqno (28)
$$
for a given value of $\Lambda$.
In Fig. 3, we display the region of parameter space
$( 4\pi G\mu , \Lambda ) $ where black holes will not form. An important
way in which our results differ from the results for gauge strings is
that circular gauge string loops of {\it any} size
and tension will collapse to form black holes whereas only
{\it large} loops of relatively massive global
string can possibly form black holes. The dependence on the size of the
loop is hidden inside the parameter $\Lambda$.

Note that, since we have ignored certain factors like the radiation
back-reaction, turbulence\refto{ps}, the gravitational radiation
and the universal expansion, we can safely say when
black holes will {\it not} form but we cannot be absolutely sure of
when black holes will form. Furthermore, we have only treated the
case of a circular loop which is most favoured to collapse to a black
hole. If the loop is not circular, black hole formation is even less
likely\refto{sh2}.

A specific value of the string tension is relevant if we consider
global strings as possible seeds for galaxy formation. Then,
$4\pi G\mu \approx 10^{-5}$ and for such strings to form black holes,
we certainly need $\Lambda > 100$. Circular loops of this size
stretch far beyond the current horizon and so we conclude that
global strings relevant for galaxy formation will not form
black holes.

\beginsection {\it{Acknowledgements:}}

We would like to thank Jaume Garriga, V. F. Mukhanov and Alex Vilenkin
for discussions. JF would also like to thank R. Calm for help with the
computing. JF was supported in part by a Fulbright Fellowship from
La Caixa and TV in part by the National Science Foundation.

\references

\item{*} Present address: Departament de Matematica Aplicada III,
Escola Universitaria Politecnica, Avda. Lluis Santalo, s/n, 17002
Girona, Spain.

\refis{sh} S. W. Hawking, Phys. Lett. B{\bf 246}, 36 (1990).

\refis{sh2} S. W. Hawking, Phys. Lett. B{\bf 231}, 237 (1989).

\refis{adjq} A. Dabholkar and J. Quashnock in ``The Formation and
Evolution of Cosmic Strings'', ed. G. W. Gibbons, S. W. Hawking
and T. Vachaspati, Cambrdige University Press, 1990.

\refis{rdps} R. L. Davis and E. P. S. Shellard, Phys. Lett. B{\bf 214},
219 (1988).

\refis{ps} C. Hagmann and P. Sikivie, Nucl. Phys. B{\bf 363}, 247 (1991).

\refis{ew} E. Witten, Phys. Lett. B {\bf 153}, 243 (1985).

\refis{avtv} A. Vilenkin and T. Vachaspati, Phys. Rev. D{\bf 4},
1138 (1987).

\refis{flregge} F. Lund and Regge, Phys. Rev. D{\bf 14}, 1524 (1976).

\refis{jackson} For example, see J. D. Jackson, ``Classical Electrodynamics'',
Wiley, 1975.

\endreferences

\vfill
\eject

\beginsection{Figure Captions}

\item{1.} The energy (per unit length) radiated from the loop, $E/2 \pi R_0$,
in units of $\eta^2$ as a function of time, $t/R_0$.

\item{2.} The function $f(t)$ versus $t/R_0$ for various values of the
parameter $\Lambda$.

\item{3.} The region of parameter space $(\Lambda , 4\pi G\mu )$ where
black holes cannot form is shown as the unhatched region. The hatched
region is where it might be possible for black holes to form.

\endjnl
\end